# Polar optical scattering in ellipsoidal nanoclusters

**Hrach S Nikoghosyan[1], Gor Nikoghosyan[2*]**

[1]Shirak State University, 4 Paruyr Sevak Str., Gyumri, 3126, Armenia
[2]Yerevan State University, 1 Alex Manoogian Str., Yerevan, 0025, Armenia
[*] Author to whom any correspondence should be addressed.
**E-mail:** nikoghosyan@ysu.am

**Abstract**

The influence of the specific geometry of three-dimensional confinement on electron-vibrational coupling in $InAs/GaAs$ nanoclusters shaped as highly oblate ellipsoids of revolution is considered. Optical phonon relaxation processes are analyzed taking into account the law of conservation of angular momentum projection and the spatial symmetry of the dimensional confinement. The conditions for the emission of chiral optical phonons carrying orbital angular momentum along the structure's growth axis are analyzed. Intraband relaxation transitions with the emission of $LO$ phonons with zero angular momentum, with characteristic anisotropy of the emission direction, leading to nonmonotonic size dependences for the $e-ph$ coupling coefficient, are considered.

## 1. Introduction

Quantum confinement in nanostructures makes it relevant to study the dependence of carrier relaxation processes on the geometric symmetry of spatial confinement, without invoking possible physical relaxation mechanisms proposed at various times [1-12]. The symmetry of the spatial confinement manifests itself in the explicit form of the envelope wave functions of the electronic and vibrational subsystems, modifying the selection rules, mechanism, and rate of quantum transitions during hot carrier relaxation. Below, a model medium with properties attractive for device applications is a nanoheterostructure of self-organized quantum dots (QDs) $InAs$ embedded in a microresonator (MR) $GaAs$ shaped as a rectangular parallelepiped with dimensions $A\times B\times C$, where polar optical scattering is dominant. Optical oscillations of the MR induce long-range electric fields that penetrate the QDs and cause intraband relaxation of excited carriers. As a result, optical phonon excitations are emitted in accordance with the selection rule based on orbital angular momentum, which are influenced by the effects of size constraints and the specific geometry of the QD-MR system. In this case, a rigid spectral position and a small width of the relaxation window are fixed, as well as a resonance of the intraband relaxation rate with an energy gap of the relaxation transition close to the energy of bulk phonons $GaAs$ [10, 11, 12]. According to experimental observations, QD $InAs$ in the environment of $GaAs$ at certain stages of growth represent disk-shaped clusters, which can be likened to highly oblate ellipsoids of revolution (HOE QD), which is important for the analytical interpretation of issues [13], [14]. The study of the influence of charge carrier confinement geometry on the course of relaxation processes is reduced to the consideration of the dimensional factors of the interaction coefficient of an electron with the modes of the phonon continuum of a microresonator (MR), taking into account the law of conservation of the projection of angular momentum during the relaxation transition. The size dependences of electron-phonon coupling in the HOE QD $InAs$ are studied at low excitation power density and low temperature $\left(k_B T \ll \hbar\omega_{LO_{\vec{q}}}\right)$. Below, the reflection of phonons from the contact boundaries of QDs and MRs is neglected, since the dispersion laws of optical phonons in the materials of nanoclusters and the matrix overlap due to the proximity of the values of elastic constants and it is necessary to use the concept of a single system of optical phonons for the entire ensemble [7], [15]. The comparatively slower relaxation of carriers with the emission of acoustic phonons is also neglected due to the low population of modes in the low-temperature process regime and the small value of the deformation potential of interaction.

## 2. Intensity and selection rules of intraband relaxation in an ellipsoidal nanocluster

The $e-ph$ interaction coefficient, which determines the rate of intraband relaxation, is represented by the following matrix element

$$g = \langle f | \hat{H}_{e-ph} | i \rangle \tag{1}$$

where $\hat{H}_{e-ph}-$ is the interaction Hamiltonian, $|i\rangle = |e, n_{\vec{q}}\rangle, |f\rangle = |g, n_{\vec{q}}+1\rangle -$ are the wave functions of the initial and final states of the system, which are, respectively, the products of the single-electron wave functions of the discrete states of the HOE QD $|e\rangle, |g\rangle$ and the wave functions of the harmonic oscillator $\Phi\left(\boldsymbol{Q}_q\right)$ in vibrational states with the number of phonons $n_{\vec{q}}, n_{\vec{q}}+1,\ n_{\vec{q}} = \left(\exp\left(\hbar\omega_{LO_{\vec{q}}}/k_B T\right)-1\right)^{-1}$. Here, the phonon mode of the resonator is modeled by a harmonic oscillator. The projections of the phonon wave vector, taking into account the dimensional constraint of localized phonon modes within the matrix $GaAs$, have the form $q_x = \frac{\pi}{A}n_1, q_y = \frac{\pi}{B}n_2, q_z = \frac{\pi}{C}n_3 -\ n_i\,(i=1,2,3) = 1,2,3,\ldots$. The stationary states $|e\rangle, |g\rangle$ are determined by the Schrödinger equation for the HOE QD model in a quasi-spherical system of orthogonal curvilinear coordinates $r, \rho, \varphi\left(0 \le \rho \le c, 0 \le r \le c, 0 \le \varphi \le 2\pi\right)$, where

$$x = nr\sin\theta\cos\varphi = n\rho\cos\varphi, y = nr\sin\theta\sin\varphi = n\rho\sin\varphi, z = r\cos\theta = r\sqrt{1-\frac{\rho^2}{r^2}} = \sqrt{r^2-\rho^2},$$

and the ellipsoid surface has the form of a sphere with radius $c = const$ inscribed in the original ellipsoid of revolution [13]. Within the framework of the approach [13], [14], in the limiting case of a strongly oblate ellipsoid of revolution with the rotation axis $z$ and semi-axes $a, c$, where $n = \frac{a}{c} \gg 1-$ is the degree of oblateness of the ellipsoid, the Schrödinger equation admits an exact solution with complete separation of variables (see Appendix 1). A simple non-degenerate zone with an energy minimum at point $\vec{K} = 0$ of the Brillouin zone is assumed. The geometry of the HOE QD imposes boundary conditions of two possible types on the wave function of the carrier, thereby predetermining the decomposition of the state of motion into two nonequivalent states: 1) the state of one-dimensional limited motion in the radial direction with energy $E(k_\Gamma)=\frac{\hbar^2 k_\Gamma^2}{2m_c^*(\Gamma)}$, satisfying the conditions of continuity of the wave function and its derivative on the surface of the sphere $r=c$, which corresponds to the main structure of the electron spectrum with large distances between levels; 2) the state of cyclic motion with respect to the variables $\rho$ and $\varphi$ with energy $E(p_\Gamma)=\frac{\hbar^2 p_\Gamma^2}{2m_c^*(\Gamma)}$, with the requirement of cyclicity of the wave function with respect to $\rho,\varphi$, forming a substructure of closely spaced levels of the HOE QD spectrum. Here $k_\Gamma, p_\Gamma -$ are the discrete components of the particle's wave vector, which, based on the boundary conditions for the $r,\rho,\varphi$ variables, determine the energy spectrum. Accordingly, the stationary size-quantized states of an electron in a rectangular potential well of a finite-depth HOE QD are represented by the following wave functions:

$$\psi_\Gamma(\vec{r})=Nu_{c0}(\vec{r})\exp(iM\varphi)J_M(p_\Gamma n\rho)\cos(k_\Gamma r) \quad (2)$$

Where $u_{c,0}(\vec{r})-$ is the periodic Bloch wave function of an electron at the center of the Brillouin zone for the $c-$zone, $J_M -$ is the Bessel function of the first kind of order $M$, and $N-$ is the normalization constant. The energy values and the relative positions of the electron levels are represented by the sum of $E_{(f,M,S)}^\Gamma=E(k_\Gamma)+E(p_\Gamma)$, where the $p_\Gamma$ values are determined by the conditions of cyclicity of the wave function with respect to the variables $\theta$ (i.e., $\rho$) and $\varphi$.

$$p_\Gamma=\frac{\Lambda_S(M)}{a}=\begin{cases}\frac{\Upsilon_S(M)}{a}, M=1,3,\ldots\\ \frac{Z_S(M)}{a}, M=2,4,\ldots\end{cases}$$

Here $\Upsilon_S(M)-$ are the dimensionless roots of the Bessel functions, which are characterized by the number $S$, and $Z_S(M)-$ are the dimensionless roots of the derivative of the Bessel functions. $k_\Gamma$ is determined from the standard dispersion equation [13]. $f-$ is the quantum number characterizing the level of the main structure, and $M,S-$ are the quantum numbers of the substructure levels. In this case, in a quantum dot of composition $InAs$ in a matrix $GaAs$, only one electron level of the main structure $f=1$ is usually accommodated. During the relaxation transition $|i\rangle\to|f\rangle(|e\rangle\to|g\rangle)$, the state of the transverse motion of the charge carrier in the $xy$ plane changes due to self-consistent interaction with the macroscopic electric field generated by ion displacements during the propagation of optical vibrations in polar crystals. Since in the $|e\rangle,|g\rangle$ states in the axially symmetric field of an ellipsoidal cluster, the projection of the orbital angular momentum of the transverse motion onto the $z$ axis (in the direction of the short-range HOE QD along the structure growth axis), $L_z=M\hbar$ is the integral of motion, the conservation law of the projection of angular momentum during the quantum transition $|i\rangle\to|f\rangle$ must be satisfied. Obviously, for the transition $|e\rangle\to|g\rangle$ between the states of the HOE QD with $M=M'$ that are degenerate with respect to the projection of the orbital angular momentum, we have

$$L_z^{|e\rangle} = L_z^{|g\rangle} = M\hbar, \tag{3}$$

where $L_z^{|e\rangle}, L_z^{|g\rangle}$ – are the projections of the momentum onto the $z$ axis, respectively, for the initial and final states of the charge carrier involved in the transition. This corresponds to the emission of $LO$ phonon excitations that do not possess orbital angular momentum, in which longitudinal displacements of the ions occur in the direction of propagation of the $LO$ oscillations. And during relaxation transitions between non-degenerate in the projection of the orbital momentum states of the spectrum of the substructure of the HOE QD with projections of the orbital momentum $L'_z = M'\hbar, L_z = M\hbar, (M' \neq M)$, the law of conservation of orbital momentum has the form

$$L'_z = L_z + L_{zph}. \tag{4}$$

In the presence of a rotational degree of freedom of collective lattice vibrations without an inversion center, the angular momentum $L'_z - L_z = L_{zph}$ can be carried away only by $TO$ phonon excitations. And this is only the case if the material for the HOE QD is a medium that has a helical axis of symmetry $z$, combining rotations at angles different from $2\pi$, with translation along the axis of rotation. $TO$ phonons can carry away orbital angular momentum in the sense that, when transitioning from $M' \neq M$, in particular, two perpendicular linear $TO$ oscillations with a phase shift of a quarter of the period $(90^\circ)$ can be formed, when added together, circularly polarized quanta of collective oscillations of the crystal lattice can be formed, possessing a quantized angular momentum $(M' - M)\hbar$, forcing atoms to move not in a straight line, but along circular trajectories. In the absence of a rotational degree of freedom of collective lattice vibrations, optical phonon relaxation during the $|e\rangle \to |g\rangle$ transition in the $M' \neq M$ case will be forbidden by the angular momentum selection rule, which speaks in favor of a slowdown in energy relaxation processes in QDs with ellipsoidal symmetry. According to recent studies [16], the emission of chiral phonons carrying angular momentum is indeed possible in media with a helical axis of symmetry. Thus, $TO$ phonons emitted during the $|i\rangle \to |j\rangle (M' \neq M)$ relaxation transition at phonon resonance and propagating along the symmetry axis of the HOE QD structure made of such a material will have a certain orbital angular momentum depending on the nature of the vibrations of the lattice atoms. In the HOE QD $InAs$ with a zinc blende lattice of symmorphic structure, which does not have a non-symmorphic symmetry element, which is the screw axis, $TO(M \neq M')$ relaxation with the emission of chiral optical phonons carrying angular momentum is forbidden by the law of conservation of angular momentum. It is obvious that by selecting the appropriate material, dimensional proportions and structural parameters of the HOE QD $a, c, I(k)$, where $I(k) = \Delta E_c - E(k_\Gamma)$ – is the energy interval of the placement of the substructure levels actually appearing in the HOE QD, $\Delta E_c$ – is the height of the potential step at the QD-matrix heterointerface, it is possible for practical purposes to increase the number of states of the substructure corresponding to a certain size-quantized level of the main structure. This will make it possible to use a system of discrete families of substructure levels to study, in particular, the processes of intraband relaxation of carriers under conditions of size quantization and anisotropy of the directions of propagation of emitted optical phonons, to create conditions for the emission of chiral phonons carrying angular momentum and opening up new prospects in spintronics, nanophotonics and biosensors.

## 3. $LO$ phonon relaxation in an ellipsoidal nanocluster $InAs/GaAs$

The Hamiltonian of the $e-ph$ interaction associated with $LO$ oscillations, which play a major role in scattering processes in polar materials, has the form $\hat{H}_{e-ph}=\sum_{\vec{q}}F_q Q_{\vec{q}}\sin(\vec{q}\vec{r})$, where $LO-$phonon modes are introduced using the continuous medium model [17] (see Appendix 2). Here $F_q=\frac{2}{\sqrt{N}}\frac{\bar{e}e^*}{\varepsilon_0 q\Omega_0}$, $q^2=q_x^2+q_y^2+q_z^2-$ is the square of the wave vector of ion oscillations, $Q_q-$ are the generalized normal coordinates corresponding to certain $q$, $\left(e^*\right)^2=\bar{M}V_0\omega_{LO_q}^2\varepsilon_0\left(\frac{1}{\chi_\infty}-\frac{1}{\chi}\right)-$ are the effective charge of the ion associated with the difference between the low- and high-frequency values of the relative permittivity, $\Omega_0-$ is the volume of the primitive cell of the crystal, $\varepsilon_0-$ is the electrical constant, $\bar{M}-$ is the reduced mass of the primitive cell, $N=\frac{V}{\Omega_0}-$ is the number of primitive cells in the crystal, $\vec{e}_{\vec{q}}-$ is the unit polarization vector, $\bar{e}-$ is the charge of the electron taking into account the sign. Taking into account the long-range nature of macroscopic fields induced by polar optical vibrations, the small change in the energy of $e-ph$ interaction within the unit cell and passing in the interaction operator to quasi-spherical coordinates $\rho,r,\varphi$, after simple transformations, for the $e-ph$ coupling coefficient (1) we obtain (see Appendix 3)

$$g_{LO_q}=-\sum_{\vec{q}}B_q I\left(q_x,q_y,q_z\right)\sqrt{n_{\vec{q}}+1}, \tag{5}$$

$$I\left(q_x,q_y,q_z\right)=2\int_0^1 dt\int_0^1 dl J_{M'}^*\left(\Lambda_{S'}(M')t\right)J_M\left(\Lambda_S(M)t\right)\cos^2\left(k_\Gamma cl\right)t\int_0^\pi e^{i(M-M')\varphi}\sin\left(\beta\sin(\varphi+\gamma)+\eta\right)d\varphi,$$

$$B_q^2=\frac{2e^2a^4c^2\hbar\omega_{LO_q}N_1^2N_2^2}{ABC\varepsilon_0 q^2}\left(\frac{1}{\chi_\infty}-\frac{1}{\chi}\right),t=\frac{\rho}{c},l=\frac{r}{c},tg\gamma=\frac{q_x}{q_y}=\frac{B}{A}\cdot\frac{n_1}{n_2},$$

$$\beta=at\sqrt{q_x^2+q_y^2},\eta=q_z c\sqrt{l^2-t^2}.$$

In the first order of perturbation theory, the rate of intraband relaxation of the electron subsystem of a quantum dot during a transition with the emission of an elementary excitation with energy $\hbar\omega_{LO_{\vec{q}}}$ is given by the following expression

$$W_{fi}=\frac{2\pi}{\hbar}\sum_{\vec{q}}\left|g_{LO_{\vec{q}}}\right|^2\delta\left(E_i-E_f+\hbar\omega_{LO_{\vec{q}}}\right)$$

Below, the analysis of the dimensional dependencies of the coupling coefficient $e-ph$ is carried out in a simplified, practically feasible limiting case

$$tg\gamma\to 0\left(A\gg B\right),\eta\to 0\left(C\gg c\right) \tag{6}$$

at values of $\hbar\omega_{LO_q}\left(GaAs\right)=35,4meV;\chi=13,2;\chi_\infty=10,9$, using the example of an intraband relaxation transition between the following states of the HOE QD degenerate with respect to the projection of the orbital angular momentum with $M=M'=2$ (Fig. 1)

$$|g\rangle = \psi_g = N_1 u_{c,\vec{k}'}(\vec{r}) e^{2i\varphi} J_2(p_{\Gamma 1} n\rho)\cos(k_\Gamma r), |e\rangle = \psi_e = N_2 u_{c,\vec{k}}(\vec{r}) e^{2i\varphi} J_2(p_{\Gamma 2} n\rho)\cos(k_\Gamma r) \quad (7)$$

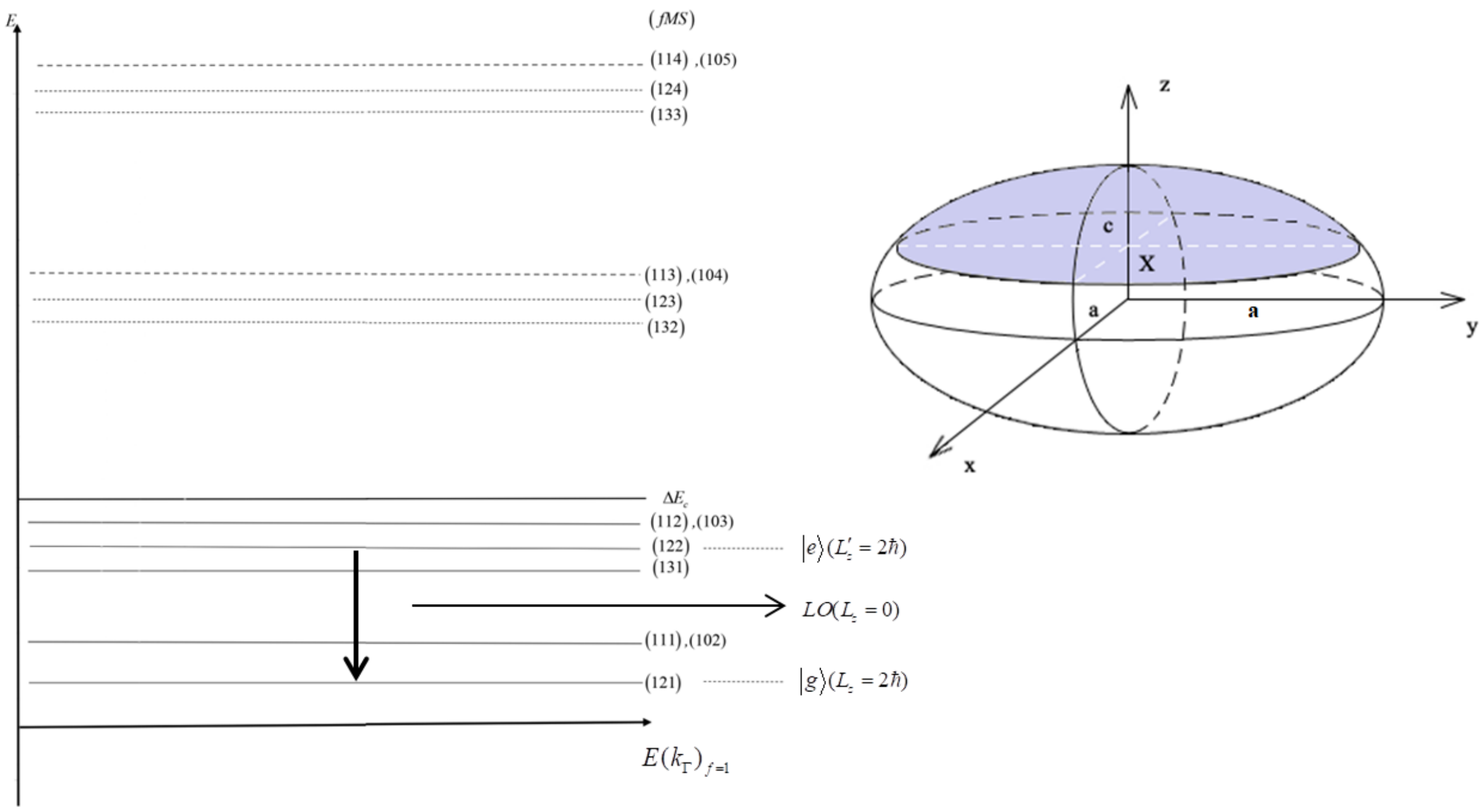

Fig. 1. Scheme of the intraband relaxation transition $|e\rangle \to |g\rangle$ at $LO$ phonon resonance between the states of the substructure of the spectrum of the HOE QD.

## 4. **Discussion of results**

The resonance of the interlevel transition $|e\rangle \to |g\rangle$ between states (7) with the energy of the $LO$ phonon, accurate to the width of the electron levels $(\sim 0{,}1 meV)$

$$E^{(e)}_{(1,2,2)} - E^{(g)}_{(1,2,1)} \approx \hbar\omega_{LO_{\vec{q}}}(GaAs)$$

is observed with the following structural parameters: $a \approx 37nm -$ is the average radius, $2c = 2{,}4nm -$ is the maximum thickness of the cluster, where $E^{(g)}_{(1,2,1)} = 532{,}3meV$, $E^{(e)}_{(1,2,2)} = 567{,}7meV$. Here $p_{\Gamma 1} = p_{(1,2,1)} = \dfrac{Z_{S=1}(M=2)}{a}, p_{\Gamma 2} = p_{(1,2,2)} = \dfrac{Z_{S=2}(M=2)}{a}$, $k_\Gamma c = 0{,}73$ [13], $Z_{S=1}(M=2) = 3{,}05424; Z_{S=2}(M=2) = 6{,}70613 -$ are the first dimensionless roots of the derivative of the Bessel functions at $M = 2$. The overlap integral $I(q_x, q_y, q_z)$ is then transformed to the form

$$I(q_x, q_y, q_z) = -2\pi \int_0^1 dt J_2^*(Z_{S=1}(M=2)t) J_2(Z_{S=2}(M=2)t) t \int_0^1 dl \mathrm{E}_0(\beta)\cos^2(kcl),$$ where

$\mathrm{E}_0(\beta) -$ is the zero-order Weber function. Figure 2 shows the dependence of the coupling coefficient $e - ph$ on the size of the MR, where the effect of phonon localization is manifested.

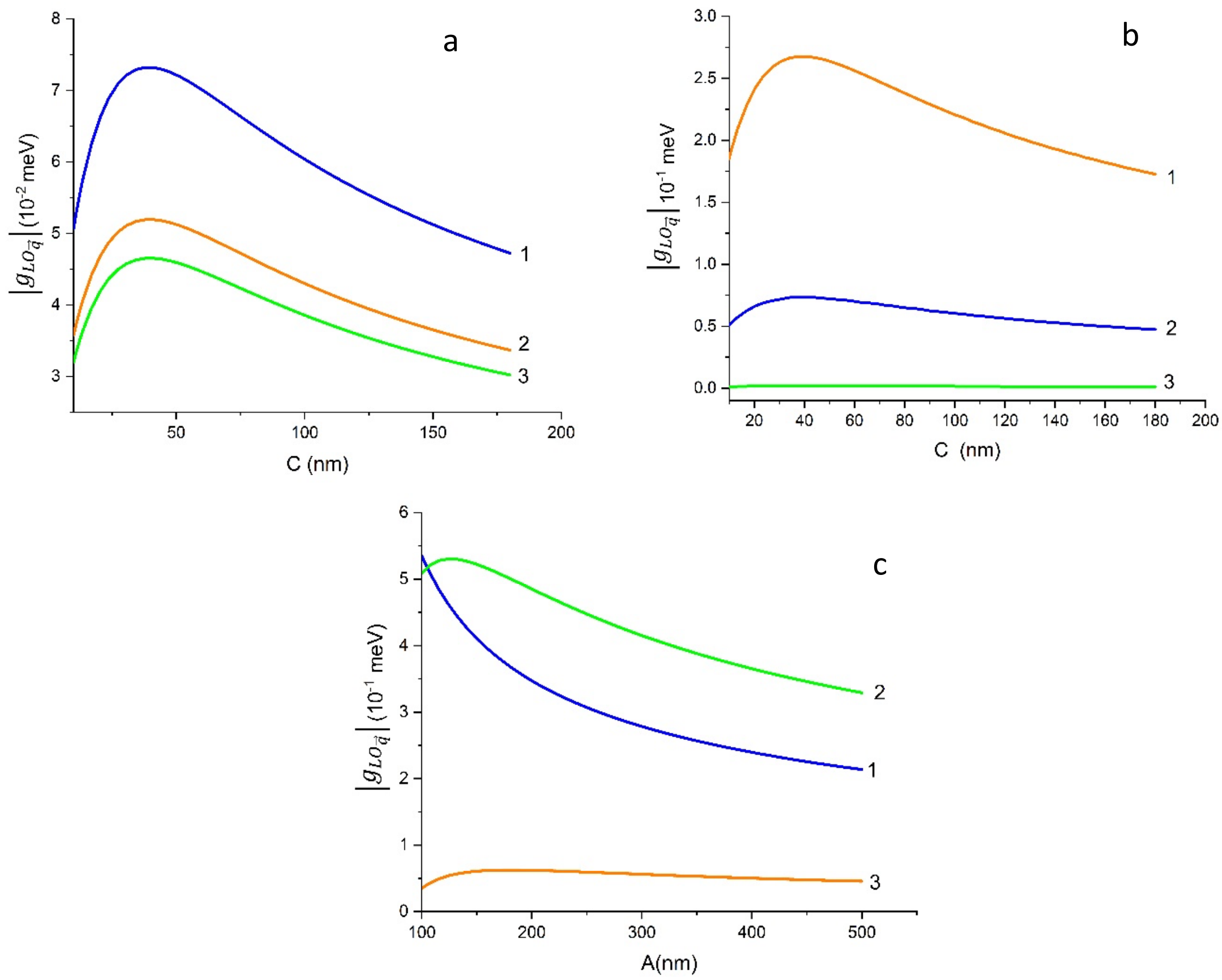

Fig. 2. Dimensional dependences of the coupling coefficient $e-ph$ $a$) on $C$ for a $(111)-$phonon, at $A=1)200;2)300;3)350(nm), B=40(nm)$; $b$) on $C$ for a $1)(222);2)(111);3)(555)-$phonon, at $A=200(nm), B=40(nm)$; $c$) on $A$ for a $(111)-$phonon, at $B=1)30;2)70;3)50(nm), C=15(nm)$.

Non-monotonic dependences of $\left|g_{LO_{\vec q}}(C)\right|$, $\left|g_{LO_{\vec q}}(A)\right|$, with the formation of maxima of $e-ph$ interaction at certain sizes of MR ($C=37,8nm, A=200nm, B=40nm$ in Fig. $2a,b$ and $A=124,3nm, B=30nm, C=15nm$ in Fig. $2c$), which determine the components and orientation of the wave vector of $LO-$oscillations $\vec q\left(\frac{\pi}{A}n_1,\frac{\pi}{B}n_2,\frac{\pi}{C}n_3\right)$, apparently indicate the existence of anisotropy of the emission directions of $LO$ phonons during relaxation transitions in the HOE QD, which reflects the spatial distribution of the overlap regions of the wave functions of the electron and phonon subsystems in the ellipsoidal nanocluster. At large quantum numbers ($n_1=n_2=n_3=5$ in Fig. $2b$), there is a decrease in the intensity of the $e-ph$ interaction. A monotonic decrease is also observed with an increase in the size of the MR $(A,B,C \gg a,c)$. And for certain dimensional proportions $\left|g_{LO_{\vec q}}\right|=0$ $(A=350nm, B=40nm, C=180nm, n_1=n_2=n_3=1)$. It should be assumed that the maxima on the curves of the $g_{LO_{\vec q}}(C,A)$ dependences are formed when, by varying the dimensional parameters of the microresonator $A,B,C$, the wavelength of phonon excitations $\lambda = 2\pi \left( \frac{\pi^2}{A^2} n_1^2 + \frac{\pi^2}{B^2} n_2^2 + \frac{\pi^2}{C^2} n_3^2 \right)^{-1/2}$ becomes comparable with the period of spatial oscillations of the envelope wave functions of charge carriers in a quantum dot. And the sharpness of the peaks allows us to judge the degree of sensitivity of the $e-ph$ coupling coefficient to dimensional variations. The obtained results highlight the theoretical possibilities of tracking relaxation processes based on the geometry of quantum confinement. The average value of the intraband $LO-$phonon relaxation rate at resonance is $1/\tau_{LO} \sim 6 \cdot 10^{11} s^{-1}$. It is close to the value of the total relaxation transition rate of a homogeneous array of QDs, which has a slight spread in shape, size and interlevel intervals, and in order of magnitude coincides with experimental estimates obtained from direct measurements of the photoluminescence kinetics of self-organized QDs at low temperatures (from 2 to 5 K) and low excitation power density (usually $< 50 W/cm^2$) [18-21].

## 5. Conclusion

This paper examines the specific features of relaxation processes in HOE QDs, conditioned by the conservation law of orbital angular momentum projection and the geometric symmetry factor of the structures under consideration. This approach leads to the formulation of conditions for the emission of $LO$ phonons with zero angular momentum and anisotropy in the propagation direction, induced during relaxation transitions between states of the HOE QD spectrum substructure, and the emission of chiral phonons along the structure growth axis, carrying angular momentum. The size dependences of the interaction coefficient of an electron with localized $LO$ phonons in a nanosized QD $InAs/GaAs$ in the form of a strongly oblate ellipsoid of revolution, placed in a microresonator with dimensions $A \times B \times C$ are analyzed. Varying the linear dimensions of the QD-MR nanoheterosystem reveals either the formation of electron-phonon coupling maxima or a monotonic behavior of the coupling coefficient under the influence of physical factors: 1) overlap of the charge carrier wave functions and phonon excitations; 2) differences in the quantum characteristics of the electron motion states in the lateral directions and in the directions of the structure's growth axis, acting as the slow and fast subsystems, respectively. Both factors demonstrate a clear dependence of relaxation processes on the geometry of the nanostructure's quantum confinement.

**Data availability statement**
No data associated in the manuscript.

**Acknowledgements** The research was supported by the Higher Education and Science Committee of MESCS RA (Research projects № 25RG-1C005 (H.S.N.) and № 23/2IRF-1C003, 24WS-1C021 (G.N))

**Conflict of Interest**

The authors declare no conflict of interest.

**Appendix 1. Separation of variables in a quasi-spherical coordinate system.**

In order to separate the variables, a quasi-spherical system of curvilinear coordinates is introduced, which becomes orthogonal at a large value of the ellipsoid oblateness parameter $n$ . From the expression for the square of the differential of the arc length, which does not contain cross products of the differentials of the curvilinear coordinates at $n \gg 1$, the Lamé coefficients are determined and the Laplacian $\nabla^2 = \frac{\partial^2}{\partial r^2} + \frac{1}{n^2}\left[\frac{1}{\rho}\frac{\partial}{\partial \rho}\left(\rho \frac{\partial}{\partial \rho}\right) + \frac{1}{\rho^2}\frac{\partial^2}{\partial \varphi^2}\right]$ is constructed, allowing for the separation of variables. Here $\rho = r\sin\theta -$ is the cylindrical radius introduced instead of the polar angle $\theta$.

**Appendix 2. Hamiltonian of the $e-ph$ interactions.**

The potential energy of interaction of an electron located at point $\vec{r}$ with a field created by bound electric charges with density $\rho_{св.}\left(\vec{R}\right) = -div\vec{P}\left(\vec{R}\right)$ has the form

$$H_{e-ph} = -\frac{1}{\varepsilon_0}\int \vec{D}\left(\vec{R}\right)\vec{P}\left(\vec{R}\right)d\vec{R},$$

where $\vec{D}(\vec{R})$ – is the induction of the electron field at point $\vec{R}$, $\vec{P}(\vec{R}) = \frac{e^* u(\vec{R})}{\Omega_0}$ – is the vector of specific polarization of the crystal, changing with frequency $\omega_{LO_q}$ with weak dispersion of optical vibrations, $u(\vec{R})$ – is the long-wave optical shift in a primitive cell located at point $\vec{R}$. The transition to generalized normal coordinates $Q_q$, $\vec{u}(\vec{R}) = \frac{1}{\sqrt{N}} \sum_{\vec{q}} \{ Q_{\vec{q}} \vec{e}_{\vec{q}} \exp(i\vec{q}\vec{R}) + \kappa.c. \}$ and integration transforms the Hamiltonian $\hat{H}_{e-ph} = -\frac{1}{\sqrt{N}} \frac{\bar{e}e^*}{\varepsilon_0 V_0} \sum_{\vec{q}} \frac{1}{q} \{ iQ_{\vec{q}} \exp(i\vec{q}\vec{r}) + \kappa.c. \}$ of the $e-ph$ interaction, which can be represented as $\hat{H}_{e-ph} = \sum_{\vec{q}} F_q Q_{\vec{q}} \sin(\vec{q}\vec{r})$, where $\vec{q}$ – is the wave vector of ion oscillations. Here, the screening caused by the rapid movement of charges that manage to follow the lattice vibrations is neglected.

**Appendix 3. $e-ph$ coupling coefficient.**

The matrix element of interaction with limited phonons $g_{LO_q}$ consists of a sum of terms, each of which is associated with a separate lattice wave and contains factors that are linear in normal coordinates, as well as an electronic part that depends on the electronic coordinate and corresponds to a given oscillation:

$$g_{LO_{\vec{q}}} = \sum_{\vec{q}} W(\vec{q}, \vec{r}) I(Q_{\vec{q}}),$$

where $W(\vec{q}, \vec{r}) = \int_V \Psi_g^* F_q \sin(\vec{q}\vec{r}) \Psi_e dV$ and $I(Q_{\vec{q}}) = \int \Phi^*(Q) Q_{\vec{q}} \Phi(Q) dQ_{\vec{q}}$.

The electronic part of the matrix element $W(\vec{q}, \vec{r})$, where integration is carried out over the volume of the ellipsoidal QD $V = N\Omega_0$, can be represented as the sum of integrals over the volumes of the crystal cells $\Omega_0$ of the QD material:

$$W(\vec{q}, \vec{r}) = \sum_N \psi_g(\vec{r}) F_q \sin(\vec{q}\vec{r}) \psi_e(\vec{r}) \int_{\Omega_0} u^*_{c,\vec{k}'}(\vec{r}) u_{c,\vec{k}}(\vec{r}) d^3 r$$

where slowly changing functions over distances of the order of $\sim \Omega_0^{1/3}$ were taken out from under the integral signs over the elementary cells $\Omega_0$. For interactions with polar optical vibrations, the resulting macroscopic fields are long-range and change little within the cell. Accordingly, the electronic part of the $\hat{H}_{e-ph}$ can also be considered approximately constant within the unit cell and taken out from under the integral signs over the cells. In addition, for states near the edges of parabolic zones we have $u_{c,\vec{k}'} = u_{c,\vec{k}+\vec{q}} \approx u_{c,\vec{k}}$. As a result, replacing the summation over $N$ with the integral over volume $V$, for the electronic part of the matrix element we obtain

$$W(q,r) = n^2 F_q N_1 N_2 \iiint J_2^*(p_1 n\rho)\cos(kr)\sin\left(q_x n\rho\cos\varphi + q_y n\rho\sin\varphi + q_z\sqrt{r^2-\rho^2}\right)\times$$
$$J_2(p_2 n\rho)\cos(kr)\rho d\rho dr d\varphi$$

where in the electron-phonon interaction operator we have moved to quasi-spherical coordinates $\rho, r, \varphi$ . $\rho = r\sin\theta -$ is the cylindrical radius. The matrix element part $I\left(Q_{\vec{q}}\right)$ has a standard form [17]:

$$I\left(Q_{\vec{q}}\right) = \left(\frac{\hbar}{2\bar{M}\omega_{LO_{\vec{q}}}}\right)^{1/2}\left(n_{\vec{q}}+1\right)^{1/2}$$

Combining the obtained expressions, we arrive at the form $g_{LO_q} = -\sum_{\vec{q}} B_q I\left(q_x, q_y, q_z\right)\sqrt{n_{\vec{q}}+1}$ .